\documentclass[pra,aps,twocolumn,floatfix,superscriptaddress]{revtex4-2}
\usepackage{amsmath}
\usepackage{amssymb}
\usepackage{graphicx}
\usepackage{color}
\usepackage{soul}
\usepackage[utf8]{inputenc}
\usepackage[T1]{fontenc}
\usepackage{hyperref}
\usepackage{array}
\usepackage{comment}
\usepackage{amsmath}
\usepackage{mathtools}
\newcommand{\abs}[1]{\left\vert #1 \right\vert } 
\usepackage{braket}

\usepackage[normalem]{ulem}

\usepackage{hyperref}
\hypersetup{colorlinks,
        linkcolor=blue,%
        citecolor=blue,%
        urlcolor=blue}

\begin{document}

\title{Interaction imbalanced spin-orbit coupled quantum droplets}

\author{Sonali Gangwar}
\affiliation{Department of Physics, Indian Institute of Technology, Guwahati 781039, Assam, India} 

\author{Rajamanickam Ravisankar}
\affiliation{Department of Physics, Zhejiang Normal University, Jinhua 321004, PR China}
\affiliation{Department of Physics, Zhejiang Institute of Photoelectronics, Jinhua 321004, PR China}

\author{S. I. Mistakidis}
\affiliation{Department of Physics, Missouri University of Science and Technology, Rolla, Missouri 65409, USA}

\author{Paulsamy Muruganandam}
\affiliation{Department of Physics, Bharathidasan University, Tiruchirappalli 620024, Tamilnadu, India}

\author{Pankaj Kumar Mishra}
\affiliation{Department of Physics, Indian Institute of Technology, Guwahati 781039, Assam, India}

\date{\today}

\begin{abstract} 
We explore the ground states and quench dynamics of spin-orbit coupled (SOC) one-dimensional two-component quantum droplets featuring intracomponent interaction imbalance.
A plethora of miscible ground state stripe and standard flat-top or Gaussian droplets is found depending on the interplay between the SOC wavenumber and interactions. 
Deformations among these states are accompanied by controllable spin population transfer.  
Upon considering a trap we identify a transition from a bound to a trapped gas many-body state, captured through a sign change of the chemical potential, which occurs at lower (larger) atom numbers for tighter traps (stronger interactions).  The droplets breathing frequency is found to increase for larger intracomponent interaction ratio or reaches a maximum at SOC wavenumbers where the transition from non-modulated  flat-top to stripe droplets exists. Dynamical droplet fragmentation occurs for abrupt changes of the Rabi-coupling, while large amplitude quenches of the SOC wavenumber trigger
spin-demixed counterpropagating untrapped droplets or in-trap out-of-phase oscillating ones. Our results offer insights into controlled spin-mixing processes in droplets and the potential excitation of magnetic bound states, opening avenues for further research in this field.

\end{abstract}

\flushbottom

\maketitle
\section{Introduction}

Attractive interacting many-body systems are historically intriguing and challenging to tackle since they are structurally unstable. Recent advances in cold atoms allow nowadays to probe this interaction regime which, under specific circumstances, may host bound states such as quantum droplets
~\cite{bottcher2020new, Luo2020}, bubbles~\cite{katsimiga2023interactions, Edmonds_dark_drops} and extreme nonlinear waves, e.g. the Peregrine soliton~\cite{Peregrine_Engels}. Droplets are highly incompressible many-body states which have already been observed in the cold atom laboratories using short-range bosonic mixtures~\cite{Cabrera2018, Cheiney2018, D'Errico2019, Semeghini2018, Ferioli2019, guo2021lee}, as well as long-range dipolar gases~\cite{Chomaz2016, Schmitt2016, chomaz2022dipolar, trautmann2018dipolar}. For short-range settings, that we are interested in this work, quantum fluctuations  customarily modelled by the Lee-Huang-Yang (LHY) correction term~\cite{properties2005eigenvalues}, provide the key mechanism to support droplet formation~\cite{Petrov2015, Luo2020}. A frequently employed framework to describe these self-bound configurations~\cite{Petrov2015,Astrakharchik2018,bougas2024stability}, including their collisions~\cite{Ferioli2019,hu2022collisional}, collective~\cite{Tylutki2020, fei2024collective} and nonlinear~\cite{li2018two,katsimiga2023solitary,chandramouli2024dispersive} excitations, refers to a system of suitably extended Gross-Pitaevskii equations (eGPEs). The latter takes into account the dimension~\cite{zin2018quantum,ilg2018dimensional,pelayo2024phases} and system~\cite{Petrov2016} dependent LHY term. {In addition to} pure bosonic systems, droplets can also exist in Bose-Fermi~\cite{BFdrops_Pelayo,rakshit2019,rakshit2019self} and spin-orbit coupled (SOC) attractively interacting  settings~\cite{cui2018spin, Gangwar2023, Gangwar2024}.     

Yet, droplets have been mainly studied by leveraging the fixed intracomponent interaction to density ratio~\cite{Petrov2015, mistakidis2023few} where the system reduces to a single-component one (symmetric droplet setting). 
The breaking of this condition where the two components are prominently distinguishable has been only very recently considered~\cite{Englezos2023, englezos2023particle, flynn2209quantum, flynn2024harmonically, kartashov2024multipole}, including also lattice~\cite{valles2024quantum} and ring~\cite{tengstrand2022droplet} geometries where more complex phases have been identified. On the other hand, the inclusion of the SOC contribution into the two-component droplet setting can bring into play additional features such as the build-up of stripe droplet configurations~\cite{Gangwar2024, Gangwar2023} and the emergence of spin transfer, both, being prohibited in the absence of Rabi-coupling~\cite{abad2013study}. Hence, the SOC parameters (wavenumber and Rabi-coupling) may be exploited to create richer two-component droplet phases and tuned to  demonstrate controllability of the ensuing unexplored two-component droplet phases. 
In this context, for instance, structural deformations from stripe to non-modulated either flat-top or Gaussian-type droplet distributions and vice versa can occur by means of varying the SOC parameters. These processes, that are absent without the SOC term are anticipated to facilitate adjustable intercomponent population transfer. 
Similarly, they can lead to transitions i) from droplet to gas states and ii) from particle imbalanced droplets to symmetric ones. 
Another interesting prospect, concerns the dynamical response of SOC droplets. 
For instance, quenching the appropriate SOC parameters, can give rise to counterpropagating or oscillating droplets or lead to their spontaneous fragmentation. 

To address these open questions, we employ a one-dimensional two-component bosonic droplet system featuring a SOC contribution and  unequal short-range intracomponent interactions. We aim to examine the highly unexplored interplay of intracomponent repulsion and SOC parameters on the generation of droplet phases, their nonequilibrium dynamics and accompanying spin-demixing processes. Our study covers systems  both in free space and under the influence of an external trap.

We find a variety of different ground state droplet structures sharing intercomponent miscibility while exhibiting stripe (non-modulated) shapes for large (small) SOC wavenumber.   
Independently of the latter, the overall droplet background deforms from Gaussian to flat-top for increasing (decreasing)  atom number (interaction)~\cite{Gangwar2024}. 
Interestingly, a gradual distortion from stripe to non-modulated Gaussian droplets and consecutively to flat-top ones occurs for increasing Rabi-coupling an outcome consisting one of the central results of this work. 
It is also explicated, for the first time, that intercomponent population imbalance can be controlled through the parametric variations of the SOC terms. 
Particularly, population imbalance is found to increase for larger interactions or SOC wavenumber, but it vanishes for increasing Rabi-coupling. 
We also showcase that an external trap enforces Gaussian stripe droplet formation of increasing size and peak density for larger atom number and, importantly, a transition from a self-bound to a trapped gas~\cite{flynn2024harmonically} reflected in the sign change of the system's chemical potential. 
The respective transition region exists at lower (larger) atom numbers for tighter traps (stronger repulsions). 

Following a quench of the trap strength, we excite the droplet's breathing motion, which is accompanied by relatively small population transfer. We demonstrate that the  frequency of this mode increases for a larger intracomponent interaction ratio and becomes maximum for SOC wavenumbers at the transition from non-modulated flat-top to stripe droplets. Such a behavior is in line with earlier findings but referring to symmetric droplets systems~\cite{Gangwar2024}. 
Moreover, sudden changes of the Rabi-coupling are shown to lead to suppressed spin-mixing and the generation of fragmented moving droplets without the trap or oscillating excited ones in its presence. 
Additionally, we find that for SOC wavenumber quenches, untrapped droplets of individual components propagate in the same (opposite) direction for small (large) post-quench values.
They also experience spin-mixing on top of an overall breathing and de-mixing while delocalizing respectively. 
Harmonically confined droplets feature 
out-of-phase (in-phase) oscillations
for larger (smaller) quench amplitudes while being immiscible (miscible).

This work is organized as follows. 
Section~\ref{sec:2} introduces the interaction imbalanced SOC droplet setting and the respective coupled eGPEs used to describe both the ground states and the dynamics. 
In Sec.~\ref{sec:3}, we analyze the ground state phases and spin transfer processes of the two-component SOC droplets both in the stripe and the non-modulated states in terms of the interactions, atom number and SOC parameters. The impact of an external trap is also discussed. 
Section~\ref{Dyn_drops} elaborates on the dynamical response of the SOC droplets after quenching either the trap frequency to induce collective breathing evolution or the SOC parameters triggering either fragmentation or a directed droplet motion and spin-mixing whose degree depends on the quench amplitude. 
In Sec.~\ref{sec:4}, we summarize our main findings and discuss future extensions of our results. 

\section{Imbalanced SOC droplet setting and equations-of-motion}
\label{sec:2}

We consider a quasi-1D two-component bosonic system in the presence of SOC characterized by 
wavenumber $k_L$, and Rabi-coupling $\Omega$ between the involved spin-$\uparrow$ and spin-$\downarrow$ states. 
The atoms in each spin state interact repulsively, i.e., $g_{\uparrow \uparrow} \neq g_{\downarrow \downarrow} >0$ and the coupling among the spin states is attractive $g_{\uparrow \downarrow}<0$. 
This condition may give rise to a SOC QD in the presence of quantum fluctuations whose effect is taken to first-order via the LHY correction term. 
The main characteristic of our setup is that the spin states are assumed to feature different intracomponent repulsions ($g_{\uparrow \uparrow} \neq g_{\downarrow \downarrow}$) which in turn result in intercomponent particle imbalance ($N_{\uparrow} \neq N_{\downarrow}$) due to Rabi-coupling.
The SOC QD is described by the 
following set of coupled eGPEs~\cite{Tononi2019,Gangwar2022,Gangwar2023, Gangwar2024, mistakidis2021formation} which in dimensionless units read 
\begin{subequations}
 \label{eq:gpsoc:2}
\begin{align}
\mathrm{i} \partial_t \psi_{\uparrow}  = & \Bigg[ -\frac{1}{2}\partial_x^2-\mathrm{i} k_{L} \partial_x + g_{\uparrow \uparrow} \vert \psi_{\uparrow} \vert ^2 - \sqrt{g_{\uparrow\uparrow}g_{\downarrow\downarrow}} \vert \psi_{\downarrow} \vert ^2 \notag \\ 
&+\dfrac{2 \sqrt{g_{\uparrow \uparrow}g_{\downarrow \downarrow}} \delta g}{(g_{\uparrow\uparrow}+g_{\downarrow\downarrow})^2}\left(g_{\downarrow\downarrow}\vert \psi_{\uparrow} \vert ^2 + \sqrt{g_{\uparrow\uparrow}g_{\downarrow\downarrow}}\vert \psi_{\downarrow} \vert ^2 \right) 
 \notag \\ 
& - \frac{g_{\uparrow \uparrow}}{\pi}\sqrt{g_{\uparrow \uparrow}\vert \psi_{\uparrow}\vert ^2 +g_{\downarrow \downarrow}\vert \psi_{\downarrow}\vert ^2}\Bigg] \psi_{\uparrow}+ \Omega \psi_{\downarrow}, \label{eq:gpsoc1-a}   \\
\mathrm{i} \partial_t \psi_{\downarrow} = & \Bigg[ -\frac{1}{2}\partial_x^2 +\mathrm{i} k_{L} \partial_x + g_{\downarrow \downarrow} \vert \psi_{\downarrow} \vert ^2 - \sqrt{g_{\uparrow\uparrow}g_{\downarrow\downarrow}} \vert \psi_{\uparrow} \vert ^2 \notag \\ 
&+ \dfrac{2 \sqrt{g_{\uparrow \uparrow}g_{\downarrow \downarrow}} \delta g}{(g_{\uparrow\uparrow}+g_{\downarrow\downarrow})^2}\left(g_{\uparrow\uparrow}\vert \psi_{\downarrow} \vert ^2 + \sqrt{g_{\uparrow\uparrow}g_{\downarrow\downarrow}}\vert \psi_{\uparrow} \vert ^2 \right) 
 \notag \\ 
&-\frac{g_{\downarrow \downarrow}}{\pi}\sqrt{g_{\uparrow \uparrow}\vert \psi_{\uparrow}\vert ^2 +g_{\downarrow \downarrow}\vert \psi_{\downarrow}\vert ^2}\Bigg] \psi_{\downarrow}+ \Omega \psi_{\uparrow}, \label{eq:effgpsoc1-b}
\end{align}%
\end{subequations}%
where $\psi_\uparrow$ ($ \psi_\downarrow$) is the 1D spin wave function. 
Also, $\delta g = g_{\uparrow \downarrow}+ \sqrt{g_{\uparrow\uparrow}g_{\downarrow\downarrow}}$ is the mean-field balance point~\cite{Petrov2016}. 
{For completeness, we remark that the above eGPEs were extracted by adding the SOC and Rabi-coupling terms in the respective interaction imbalanced two-component 1D  eGPEs~\cite{Tononi2019,mistakidis2021formation}. However, the strict evaluation of the respective eGPEs and in particular of  the LHY term including SOC should follow the prescription of Ref.~\cite{Petrov2016} being an interesting perspective on its own but beyond the scope of this work.}
Since below the system fulfills either $N_{\uparrow} \neq N_{\downarrow}$  or $g_{\uparrow \uparrow} \neq g_{\downarrow \downarrow}$ the two-components should behave differently and we aim to probe their deviations when SOC is present. 
For convenience, we use the normalization condition~\cite{Astrakharchik2018} on the system wave function $\int\limits_{-\infty}^{\infty} 
\left[\lvert \psi_\uparrow \rvert^2 + \lvert \psi_\downarrow \rvert^2 \, \right] dx = N$. Here, the constant $N$ specifies the normalized total number of particles in the droplet with the population of the individual components given by $N_{\uparrow,\downarrow}=\int_{-\infty}^{\infty} \lvert \psi_{\uparrow,\downarrow} \rvert^2dx$. {Note that the normalized atom number $N$ can be a non-integer number, however, the total number of particles as considered in the experiment can be expressed through $\mathcal {N}$ and refers, for instance, to multiples of $10^2, 10^3, \dots$~\cite{Gangwar2024}. Therefore, if $10^4$ corresponds to the experimentally considered  particle number then $\mathcal{N}=10^4N$}.   

The energy unit of the system is expressed with respect to $\hbar\omega_\perp$, where $\omega_\perp$ is the transverse trap frequency. 
Accordingly, the time and length scales are measured in units of $\omega_{\perp}^{-1}$ and $a_{\perp}=\sqrt{\hbar/(m\omega_{\perp})}$ respectively. 
The effective 1D intracomponent interaction strengths $g_{\uparrow \uparrow}$ = $2 \mathcal{N_{\uparrow}} a_{\uparrow \uparrow}/ a_{\perp}$ and  $g_{\downarrow \downarrow}$ = $2 \mathcal{N_{\downarrow}} a_{\downarrow \downarrow}/ a_{\perp}$, while the intercomponent one is $g_{\uparrow\downarrow} = 2 \mathcal{N}_{\uparrow} a_{\uparrow \downarrow} / a_{\perp}$ with $\mathcal{N}_{\uparrow,\downarrow}$ being  the particle number in each component. 
In these expressions, ($a_{\uparrow \uparrow}$, $a_{\downarrow \downarrow}$) is the set of intracomponent 3D s-wave scattering lengths and $a_{\uparrow \downarrow}$ signifies the intercomponent one. These can be tuned in the experiment through Feshbach resonances~\cite{Semeghini2018}.  
The 1D wave function is rescaled as $ \psi_{\uparrow, \downarrow} = \tilde{\psi}_{\uparrow, \downarrow} \sqrt{a_{\perp}}$, while the SOC wavenumber $k_L \to \tilde{k}_L a_{\perp}$ and intensity (Rabi-coupling) $\Omega \to \tilde{{\Omega}}/\omega_{\perp}$. 
All tilde quantities refer to dimensionful ones. 

In a corresponding experiment the two spin states can be emulated, for instance, with the hyperfine states $\ket{\uparrow} \equiv \ket{F = 1, m_F = -1}$ and $\ket{\downarrow} \equiv \ket{F = 1, m_F = 0}$ of $^{39}$K~\cite{Cabrera2018, Cheiney2018, Ferioli2019}, while considering a total number of atoms $\mathcal{N}\sim [10^4,10^6]$. 
The quasi-1D geometry is achieved via a tight confinement of frequency $\omega_\perp = 2\pi \times 600$Hz in the transverse directions, which leads to $a_\perp = 0.657\,\mu$m. Also, across the elongated direction, a box potential using a digital micromirror device can be used~\cite{navon_quantum_2021}.  
Moreover, the intervals of 3D scattering lengths $a_{\uparrow \uparrow} =2/N[0.1242a_0, 0.3104a_0]$ and $ a_{\downarrow \downarrow} = 2/N[0.1242a_0, 2.4831a_0]$, where $a_0$ is the Bohr radius, can be used to reach the effective couplings $g_{\uparrow\uparrow}=[0.2, 0.5]$ and $g_{\downarrow\downarrow}=[0.2, 4]$ allowing for ratios $g_{\downarrow\downarrow}/g_{\uparrow\uparrow}=[1,20]$.
Along the same lines, $a_{\uparrow \downarrow} = 2/N [-0.0621a_0, -0.8070a_0]$ in order to have $g_{\uparrow\downarrow} = [-0.1, -1.3]$ for $\delta g = 0.1$. The dimensionless Rabi-coupling frequency interval $\Omega = [0,30]$ can be realized by tuning the Raman laser intensity~\cite{lin2011spin,zhang2012collective} in the range 
$\tilde{\Omega} = 2\pi \times [0-18 ]$kHz, and the SOC wavenumber interval $k_L = [0.1, 12]$ corresponds to variations of the laser wavelength~\cite{lin2011spin, hou2018momentum, qu2013observation} from $\lambda_L = 29.19~\mu$m to $243.25$nm~\cite{lin2011spin}. {At this point, it is important to note that the aforementioned variation in $\lambda_L$ is related to the range of the non-dimensional $k_L$ values that we are considering for a specific transverse trap frequency. The latter can be flexibly adjusted in corresponding quasi-1D experiments~\cite{mistakidis2023few}. 
This in turn implies that the actual range of the dimensionful SOC wavenumbers can be accordingly tuned by means of changing the transverse trap strength (and hence $a_{\perp}$) and the angle between the Raman lasers to match the non-dimensional $k_L$ parameter used in our simulations.}

To numerically obtain the ground state of the interaction imbalanced SOC droplets described by the time-independent version of Eqs.~(\ref{eq:gpsoc1-a}) and (\ref{eq:effgpsoc1-b}), we use the imaginary time propagation method exploiting a split-step Crank-Nicolson scheme~\cite{Muruganandam2009, Young2016, Ravisankar2021a}. On the other hand, to investigate the nonequilibrium dynamics of the SOC droplets, we rely on the real-time propagation method. In dimensionless units, the utilized box size has length $L=307$ and features spatial resolution $dx = 0.025$, while the time step of the integrator is fixed to $dt = 10^{-5}$. 
{We remark that by increasing the spatial resolution, i.e. the number of grid points, in our simulations, the observables to be presented below remain un-altered. This ensures the numerical convergence of our results both for the ground state and the dynamics.} 
For all the simulations to be presented below, we employ a Gaussian initial ansatz for each component, simultaneously imposing the anti-symmetric condition, i.e., ${\psi}_{\uparrow}(x)=-{\psi}_{\downarrow}(-x)$ on the components which facilitates the numerical identification of the SOC droplet ground state~\cite{sarkar2024signature}.

\section{Ground states of the imbalanced SOC droplets}
\label{sec:3}

A pertinent question concerns the structure of the droplet configurations and their control by varying relevant system parameters. Recently, it has been demonstrated that an imbalanced bosonic mixture in the droplet regime (where the components preserve their atom number) may arrange in miscible flat-top or Gaussian type QD distributions with distorted tails upon tuning the intracomponent interactions~\cite{flynn2024harmonically, Englezos2023}. However, in the presence of SOC, particle exchange can occur due to the Rabi-coupling and interestingly, the QDs can also assemble in the so-called stripe configurations. 
Below, we exploit different intracomponent interaction imbalances in order to characterize various ground-state QD configurations featuring intercomponent atom exchange.

\subsection{SOC droplets in free space}
\label{sec:3a}
To understand the impact of the intracomponent interaction imbalance quantified by the ratio $g_{\downarrow\downarrow}/g_{\uparrow\uparrow}$ on the SOC QD structures without a trap, we initially employ the densities $\abs{\psi_{\uparrow}}^2$ and $\abs{\psi_{\downarrow}}^2$ of the individual  components. 
\begin{figure}[!ht]
\centering\includegraphics[width=0.99\linewidth]{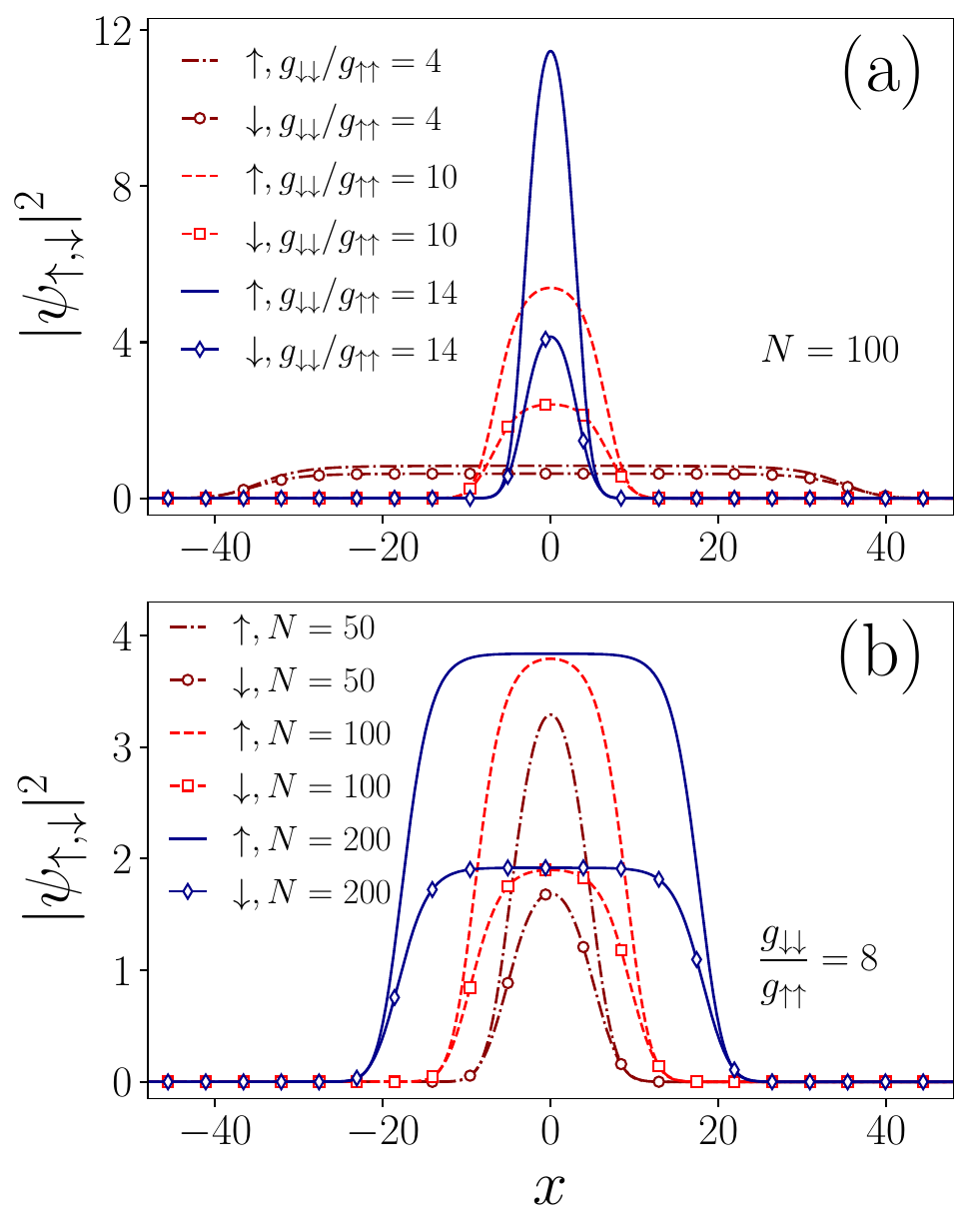}
\caption{Ground state densities of the spin-up (down) component of the SOC QD (a) for different intracomponent interaction ratio $g_{\downarrow \downarrow}/g_{\uparrow \uparrow}$ keeping the total atom number $N=100$ and (b) for varying $N$ and fixed $g_{\downarrow \downarrow}/g_{\uparrow \uparrow}=8$. Each component droplet densities transit from flat-top to Gaussian profiles upon increasing $g_{\downarrow \downarrow}/g_{\uparrow \uparrow}$ or decreasing $N$. The droplets among the components are miscible, and the distribution of the stronger interacting spin-down one becomes wider for larger (smaller) $g_{\downarrow \downarrow}/g_{\uparrow \uparrow}$ ($N$). 
In all cases, the SOC QD is characterized by $\Omega=0.5$, $k_L=0.5$, $\delta g=0.1$ and $g_{\uparrow \uparrow}=0.2$.}
\label{fig:denqdpwProfile}
\end{figure}
We first focus on SOC parameters $\Omega=0.5$, $k_L=0.5$ implying the absence of spatially modulated stripe phases since $k_L^2<\Omega${, a condition that was explicated e.g. in Refs.~\cite{Gangwar2022, Gangwar2024, Tononi2019, lin2011spin}.} Figure~\ref{fig:denqdpwProfile}(a) presents the component density profiles for different $g_{\downarrow\downarrow}/g_{\uparrow\uparrow}$, while the remaining system parameters ($N=100$, $\delta g=0.1$) including the SOC ones are held fixed. 
As it can be seen, upon increasing $g_{\downarrow\downarrow}/g_{\uparrow\uparrow}$ there are two important structural modifications in the QD distributions. 
Namely, a gradual transition from flat-top to Gaussian type miscible droplets takes places. 
This is understood by the fact that a increase in $g_{\downarrow\downarrow}/g_{\uparrow\uparrow}$ implies a stronger intercomponent attraction for maintaining $\delta g$ constant. 
This stronger intercomponent attraction is responsible for the spatial localization of the droplets and hence their deformation to Gaussian type configurations which has also been discussed in two-component droplets without SOC~\cite{Englezos2023}. 
Additionally, to identify the interaction threshold value beyond which the aforementioned transition occurs, we calculate the total kinetic energy of the droplet  as a function $g_{\downarrow\downarrow}/g_{\uparrow\uparrow}$ as depicted in Fig.~\ref{fig:norm}(a).
The sharp increase of the kinetic energy beyond $g_{\uparrow\uparrow}/g_{\downarrow\downarrow}\sim8$ signifies the transition from the flat-top to the Gaussian droplet. Indeed, flat-top droplets being highly incompressible should {minimize} their kinetic energy~\cite{Petrov2016}.  

\begin{figure}[!htp]
\centering\includegraphics[width=0.99\linewidth]{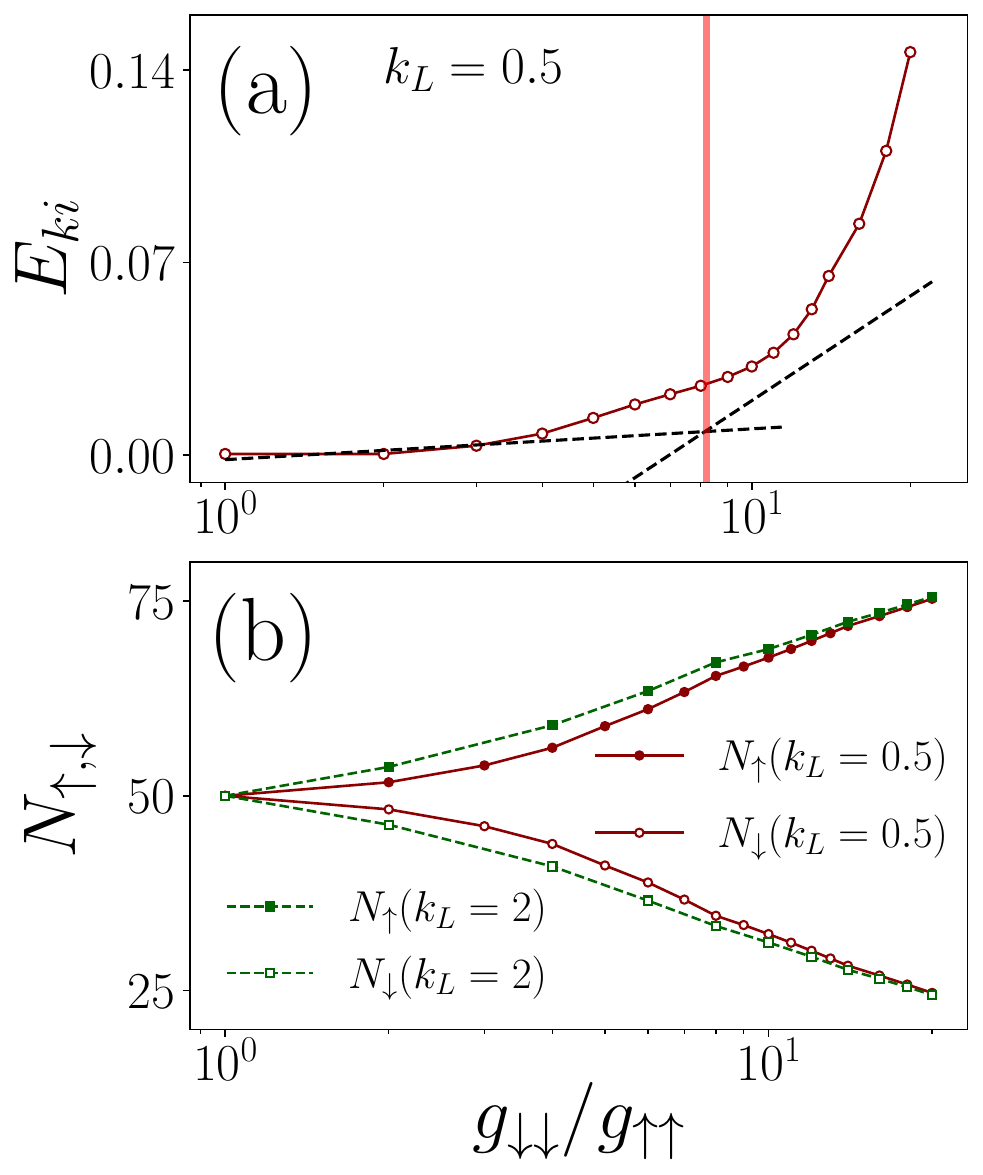}
\caption{(a) Total kinetic energy, $E_{\textrm{ki}}$, with respect to $g_{\downarrow\downarrow}/g_{\uparrow\uparrow}$ and SOC characteristics $k_L=0.5$ and $\Omega=0.5$. 
The increasing tendency of $E_{\textrm{ki}}$ with a finite rate for larger $g_{\downarrow\downarrow}/g_{\uparrow\uparrow}$ indicates the deformation of the droplets to a Gaussian shape. 
{The dotted lines provide a guide to the eye for quantifying the intersection of the interaction regions where $E_{\textrm{ki}}$ becomes almost zero (around $g_{\downarrow\downarrow}/g_{\uparrow\uparrow} \approx 1$) and the one where  $E_{\textrm{ki}}$ starts to increase (at $g_{\downarrow\downarrow}/g_{\uparrow\uparrow}>>1$).} 
(b) Variation of each component atom number as a function of   $g_{\downarrow \downarrow}/g_{\uparrow \uparrow}$ for different  $k_L$ (see legend). 
Population transfer occurs from the strongly interacting spin-down to the weakly spin-up components and it is enhanced for larger $g_{\downarrow \downarrow}/g_{\uparrow \uparrow}$. 
The remaining system parameters are the same with Fig.~\ref{fig:denqdpwProfile}.}
\label{fig:norm}
\end{figure}

Another observation is that both components having droplet character remain miscible independently of $g_{\downarrow\downarrow}/g_{\uparrow\uparrow}$ with the spin-down featuring the stronger repulsion to be wider. 
Notice here that suppressing $g_{\downarrow\downarrow}/g_{\uparrow\uparrow}$ leads to almost identical droplets in both components, thus recovering the symmetric droplet case which has been mainly explored so far~\cite{mistakidis2023few, Luo2020}. 
Importantly, a larger $g_{\downarrow\downarrow}/g_{\uparrow\uparrow}$ yields transfer of atoms among the spin components which can also be testified by monitoring the component populations as a function of $g_{\downarrow\downarrow}/g_{\uparrow\uparrow}$ shown in Fig.~\ref{fig:norm}(b). 
Indeed, atoms migrate from the strongly interacting component towards the weaker interacting one, and the particle imbalance is enhanced for increasing $g_{\downarrow\downarrow}/g_{\uparrow\uparrow}$.

For completeness, we illustrate the effect of the total atom number $N$ on the QD ground states in Fig.~\ref{fig:denqdpwProfile}(b) while fixing all other parameters and considering $g_{\downarrow\downarrow}/g_{\uparrow\uparrow}\sim 8$. 
In all cases, we observe that the droplet components assemble in a miscible fashion, with the strongly interacting spin-down component being wider and accommodating a smaller number of atoms [see Fig.~\ref{fig:norm}(a)]. 
Moreover, for larger $N$, the QD becomes broader both in amplitude and width and transforms from a Gaussian shape (e.g. $N=50$) to a flat-top (e.g. $N\gtrsim 100$) one. 
We remark that such a transition with respect to the atom number has also been discussed in the context of single-component droplets~\cite{Gangwar2024}.

\begin{figure}[!htp]
\centering\includegraphics[width=0.99\linewidth]{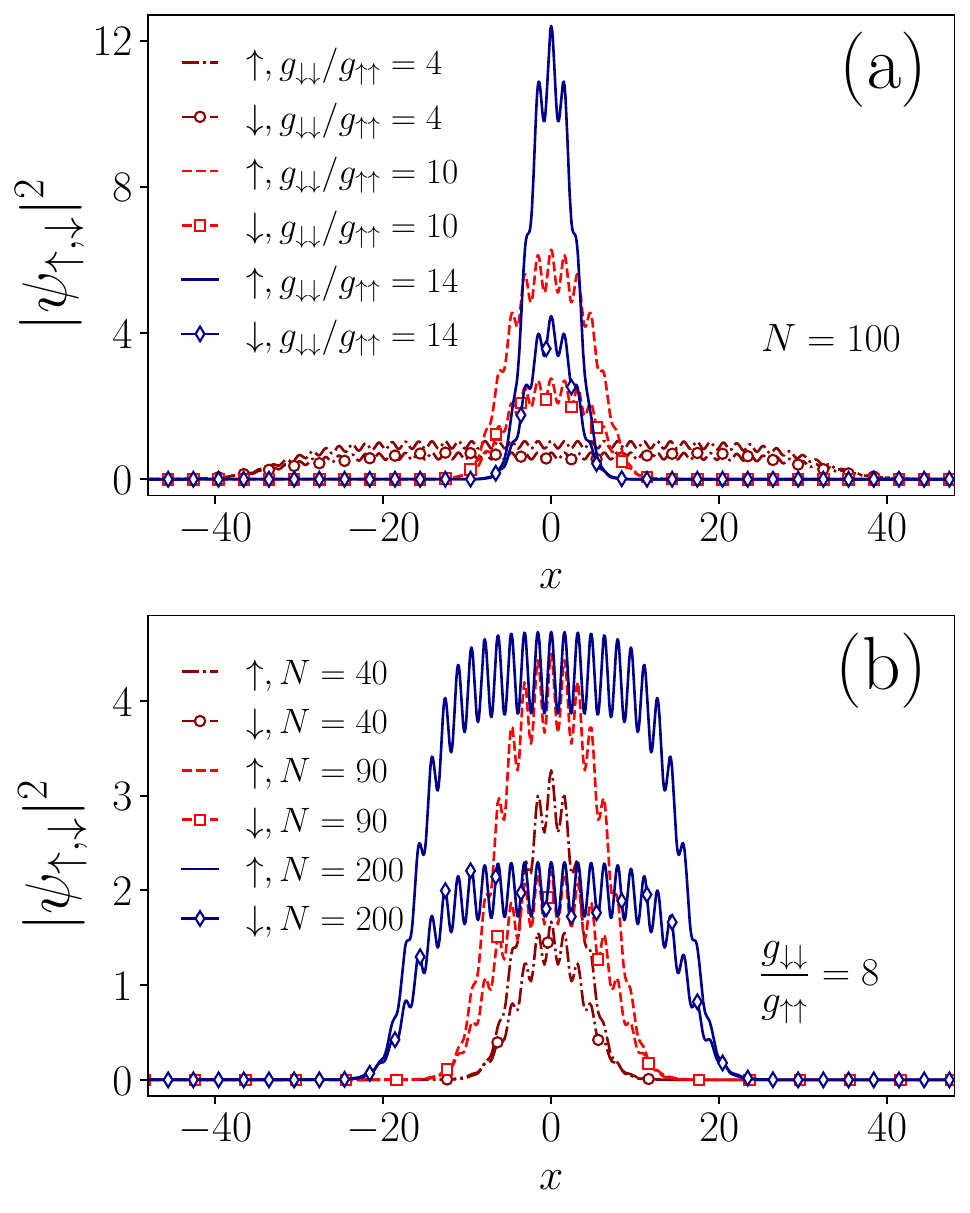}
\caption{Densities of the ground state stripe droplets of the spin-up and spin-down components (a) in terms of $g_{\downarrow \downarrow}/g_{\uparrow \uparrow}$ while considering $N=100$ and (b) for different $N$ having $g_{\downarrow \downarrow}/g_{\uparrow \uparrow}=8$. 
As it can be seen, a decreasing (increasing) $g_{\downarrow \downarrow}/g_{\uparrow \uparrow}$ ($N$) leads to the formation of stripe flat-top SOC droplets in each of the components with the latter maintaining their miscible character and hosting a larger number of stripes.  
The stripe SOC droplets are prepared with $\Omega=0.5$, $k_L=2$ $\delta g=0.1$, and $g_{\uparrow \uparrow}=0.2$.}
\label{fig:denqdsw}
\end{figure}

Similar ground state QD configurations exist also when the SOC parameters become more important, i.e., for $k_L^2>\Omega$, where stripe droplets build up in both components. 
To demonstrate this situation we employ a SOC wavenumber $k_L=2$ and Rabi-coupling $\Omega=0.5$, while investigating variations of the interaction ratio $g_{\downarrow\downarrow}/g_{\uparrow\uparrow}$ [Fig.~\ref{fig:denqdsw}(a)] and the total atom number [Fig.~\ref{fig:denqdsw}(b)]. 
The other system parameters remain fixed. 
It becomes evident that either decreasing $g_{\downarrow\downarrow}/g_{\uparrow\uparrow}$ or increasing $N$ a deformation from a Gaussian-type stripe droplet to a flat-top one occurs. 
{Naturally, flat-top droplets can host a larger number of stripes since they are substantially elongated, as was also reported in the respective interaction balance setting~\cite{Gangwar2024}.} 
As in the $k_L=0.5$ case, the weakly interacting component has relatively smaller width and larger peak density.
Finally, the transfer of atoms takes place from the strongly interacting spin-down to the weakly interacting spin-up component, see Fig.\ref{fig:norm}(b). 
This population transfer is almost the same both in magnitude and trend with the $k_L=0.5$ scenario.

\begin{figure}[!htp]
\centering\includegraphics[width=0.99\linewidth]{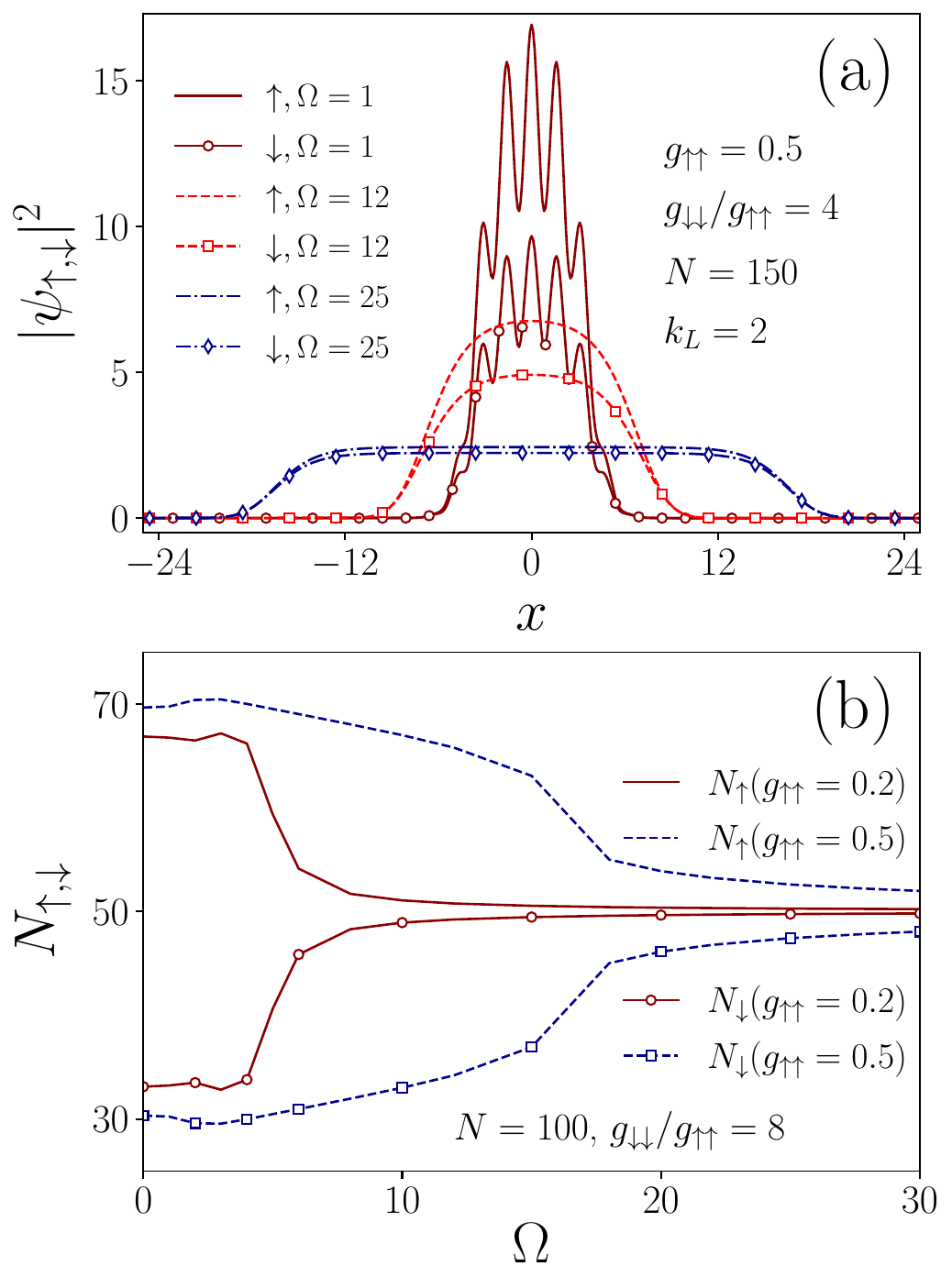}
\caption{Ground state densities of the spin-up and spin-down  droplet components for different Rabi-couplings $\Omega$ (see legend) and fixed $g_{\downarrow \downarrow}/g_{\uparrow \uparrow}=4$, $N=150$, and $k_L=2$. An increasing $\Omega$ results in the deformation from stripe to standard and afterwards  wider droplets with more pronounced flat-top and almost equal intercomponent population. (b) The population of each spin component as a function of $\Omega$ for different intracomponent interactions $g_{\uparrow\uparrow}=0.2$ and $g_{\uparrow\uparrow}=0.5$ with fixed $N=100$, $k_L=2$ and $g_{\downarrow\downarrow}/g_{\uparrow\uparrow}=8$. 
The $\Omega$ threshold above which the components become almost equally populated depends on the interactions.}
\label{fig:denqdpwithome}
\end{figure}

As a next step, we analyze the impact of the Rabi-coupling on the shape and population imbalance of the two-component droplet. 
Characteristic density profiles of both droplet components are depicted in Fig.~\ref{fig:denqdpwithome}(a) with respect to the Rabi coupling. 
It can be seen that a larger $\Omega$ results in the structural modification from the stripe to standard droplets close to the threshold $k_L^2=\Omega$. 
Further increasing $\Omega$ enforces a broadening of the non-modulated droplet distributions towards flat-top profiles and simultaneous suppression of the intercomponent population imbalance. 
Such density deformations upon Rabi-coupling variations have also been reported in single-component droplets~\cite{Gangwar2024}. 
Hence, an increasing $\Omega$ tends to balance the population among the components, which is attributed to the dominance of the Rabi-coupling energy term as compared to the kinetic, 
mean-field and LHY energy contributions in the droplet. 
{It is interesting to notice that other phases, such as the plane-wave and the zero-momentum ones, which have been reported in the context of SOC gases~\cite{Li_SOC_ZMphase} have not been identified in the present setup. 
Such states could be unveiled in the respective excitation spectrum by considering, for instance, non-zero quasi-momenta. 
Such a study worths to be pursued in future works.} 

Moreover, we find that the intercomponent atom transfer from the majority to the minority component is minimized for high $\Omega$~\cite{Deng:2021}. 
This can be directly inferred from the spin populations shown in Fig.~\ref{fig:denqdpwithome}(b) in terms of $\Omega$ for different interactions and $k_L=2$. 
The spin populations remain nearly constant for low $\Omega$, namely before the transition from stripe to standard configurations occurs. 
Afterwards, they start to be significantly modified and become almost the same for high enough $\Omega$. 
This process depends on the intracomponent interactions and in particular the $\Omega$ interval where the spin imbalance is enhanced is larger for stronger interactions. 
{This feature is in accordance to the behaviour of a spin-1/2 repulsive gas~\cite{abad2013study,mistakidis2022inducing} where an enhanced Rabi-coupling favors a superposition of spin-balanced components.} 
Therefore, the interactions can be used as a knob to control the spin imbalance. 
{Recall also that such spin transfer processes and in general imbalanced mixed droplet phases are absent for corresponding intracomponent interaction symmetric settings as was shown e.g. in Refs.~\cite{Gangwar2022,Gangwar2023,Gangwar2024}.}
It is worth noting that a similar behaviour of the spin populations takes place also for other values of the SOC wavenumber $k_L$ and especially the imbalance at relatively small $\Omega$ is more prominent for increasing $k_{L}$ (not shown for brevity). 

\subsection{Impact of a harmonic trap: transition from droplet to gas}
\label{sec:3b}
\begin{figure}[!ht]
\centering\includegraphics[width=0.99\linewidth]{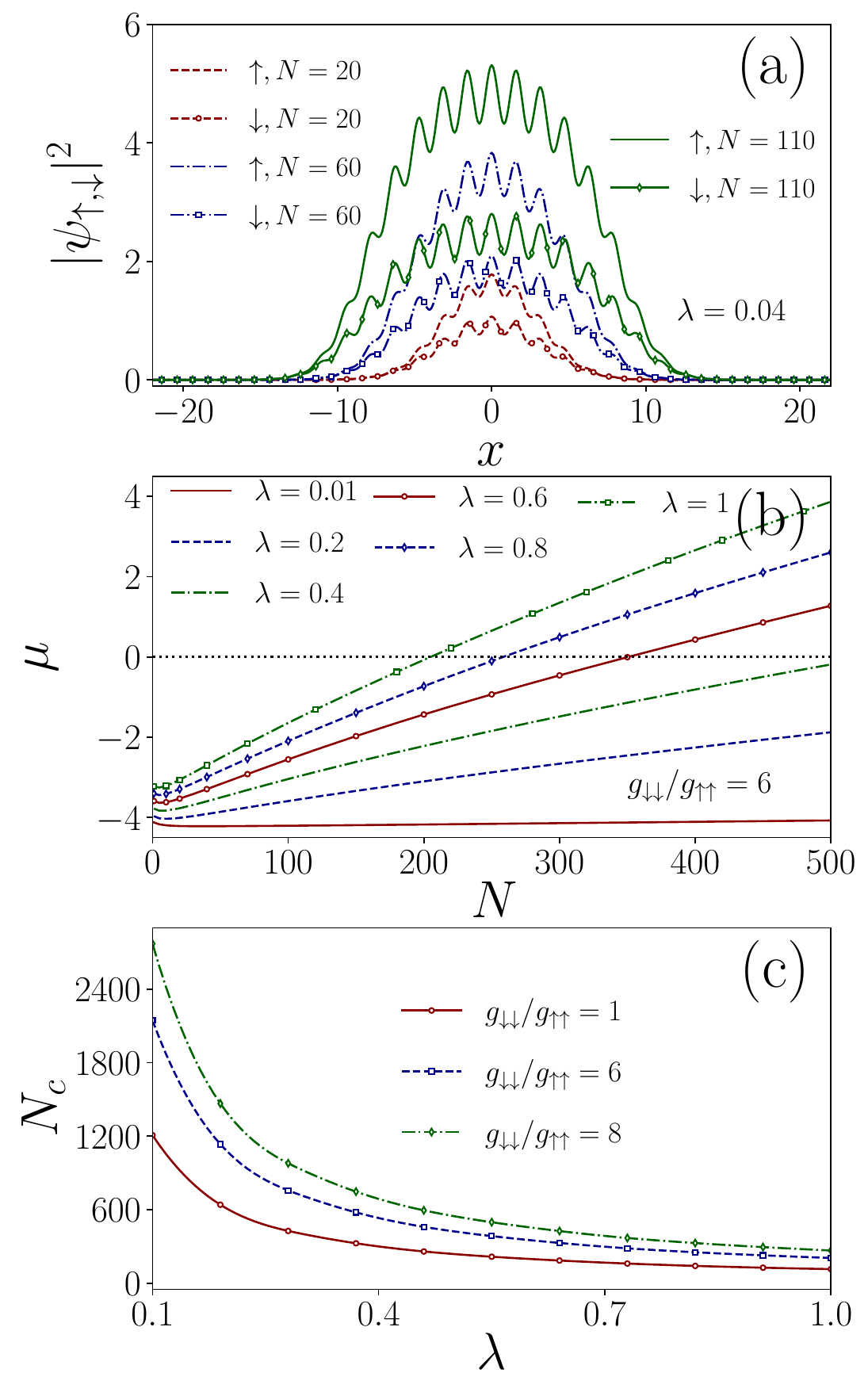}
\caption{(a) Ground state density  profile of the droplet components for different $N$ (see legends) in the presence of a harmonic trap with strength $\lambda=0.04$ and $g_{\downarrow \downarrow}/g_{\uparrow \uparrow}=6$. The stripe distributions of both components become wider for larger, $N$ while flat-top signatures are suppressed due to the trap. 
(b) Total chemical potential, $\mu$, of the binary stripe droplet as a function of $N$ for different trap strengths (see legend) when  $g_{\downarrow\downarrow}/g_{\uparrow\uparrow}=6$. A transition from a negative to positive system's energy exists for increasing $N$ becoming more prominent for tighter confinement.
(c) Critical atom number, $N_c$ for the droplet to gas transition with respect to $\lambda$ and various $g_{\downarrow\downarrow}/g_{\uparrow\uparrow}$ (see legend). 
The transition takes place at larger $N_c$ for increasing $g_{\downarrow\downarrow}/g_{\uparrow\uparrow}$ and fixed $\lambda$. 
In all cases the SOC parameters $\Omega=0.5$, and $k_L=2$, while the interaction strengths $g_{\uparrow \uparrow}=0.2$ and $\delta g=0.1$. 
}
\label{fig:denqdpwWithTrap}
\end{figure}

Having explicated the effect of the intracomponent interaction imbalance, the atom number and the Rabi-coupling on the two-component SOC droplet configurations in free space, we move to understand the impact of an external harmonic trap on the ensuing ground states. 
Specifically the eGPE Eqs.~(\ref{eq:gpsoc1-a}), (\ref{eq:effgpsoc1-b}) contain the additional $V(x)\psi_{\uparrow, \downarrow}$ term with the same trap $V(x)= (1/2)M \lambda^2 x^2$ (where $m_{\uparrow}=m_{\downarrow} \equiv M$) acting on both spin states. 
This external confinement imposes the harmonic oscillator length, $a_{\textrm{ho}}=\sqrt{\hbar/(m\lambda)}$, across the axial direction as another relevant length scale which certainly delays the formation of flat-top structures as it has already been reported for single-component droplets~\cite{Englezos2023}. 
However, the presence of an external trap is common in corresponding experimental settings~\cite{Semeghini2018,Cheiney2018,D'Errico2019}.

Density profiles of ground state SOC droplets are presented in Fig.~\ref{fig:denqdpwWithTrap}(a) for varying atom number and an exemplary relatively weak trap frequency $\lambda=0.04$, while $g_{\downarrow\downarrow}/g_{\uparrow\uparrow}=6$, $\Omega=0.5$, and $k_L=2$ are kept fixed. 
It can be readily seen that the finite trap acts against the formation of flat-top droplets for increasing atom number. Notice that for the same parameters but without the trap, both components feature a transition towards a flat-top with stripe shape as $N$ increases (not shown for brevity) [see Fig.~\ref{fig:denqdpwProfile}(b)].  
Instead, here, for larger $N$, the stripe droplet density becomes wider, featuring a striped Gaussian shape whose peak value increases  [Fig.~\ref{fig:denqdpwWithTrap}(a)]. 
The striped droplet components remain miscible, and the stronger interacting component has a slightly larger width and lower peak density. 
As expected, upon considering a tighter trap, the Gaussian profile of the striped droplet becomes more pronounced (not shown), a behaviour that remains consistent for larger atom number.

To shed further light on the role of the external trapping potential for the droplet configurations and their bound character, we subsequently calculate the total chemical potential ($\mu$) of the two-component system for varying atom number. 
As an example, we employ an interaction ratio of $g_{\downarrow\downarrow}/g_{\uparrow\uparrow}=6$ illustrated in Fig.~\ref{fig:denqdpwWithTrap}(b) for different values of $\lambda$. 
In the case of $\lambda=0.01$, $\mu $ initially (namely for quite small atom numbers, $N<20$) exhibits a decreasing tendency (i.e. it becomes more negative) for increasing $N$, and eventually after reaching a certain atom number it saturates.
This behavior indicates that the many-body state maintains its self-bound droplet character throughout the $N$ variation. 
On the other hand, as the confinement becomes tighter (namely for larger $\lambda$) the aforementioned chemical potential trend changes drastically. 
Specifically, $\mu$ starts to decrease for relatively small atom number acquiring a minimum and afterwards features a clean increasing tendency which is more pronounced for larger $\lambda$. 
However, irrespectively of $N$, it holds that $\mu<0$ and hence the system maintains its self-bound nature as long as $\lambda$ is not noticeable, e.g. $\lambda \leq 0.2$. 
The chemical potential minimum with respect to $N$, is a consequence of the balance of the kinetic, the mean-field, and the beyond mean-field LHY energy contributions at the specific $N$ depending on $\lambda$.   
We also observe that this minimum value of $\mu$ shifts towards lower $N$ for larger $\lambda$ due to the increasing influence of the potential energy. 

Interestingly, there is a critical value of the trap strength e.g. $\lambda \geq 0.4$ in our setting at which $\mu$ crosses from negative to positive values, see also the green dash-dotted line in Fig.~\ref{fig:denqdpwWithTrap}(b). 
This suggests a transition of the two-component bosonic system from a self-bound droplet phase ($\mu<0$) towards a trapped gas where $\mu>0$~\cite{du2023ground,flynn2024harmonically}. 
It is also evident that this transition occurs at smaller $N$ for increasing $\lambda$ since the energy of the trap becomes more prominent. 

\begin{figure}[!htp]
\centering\includegraphics[width=0.99\linewidth]{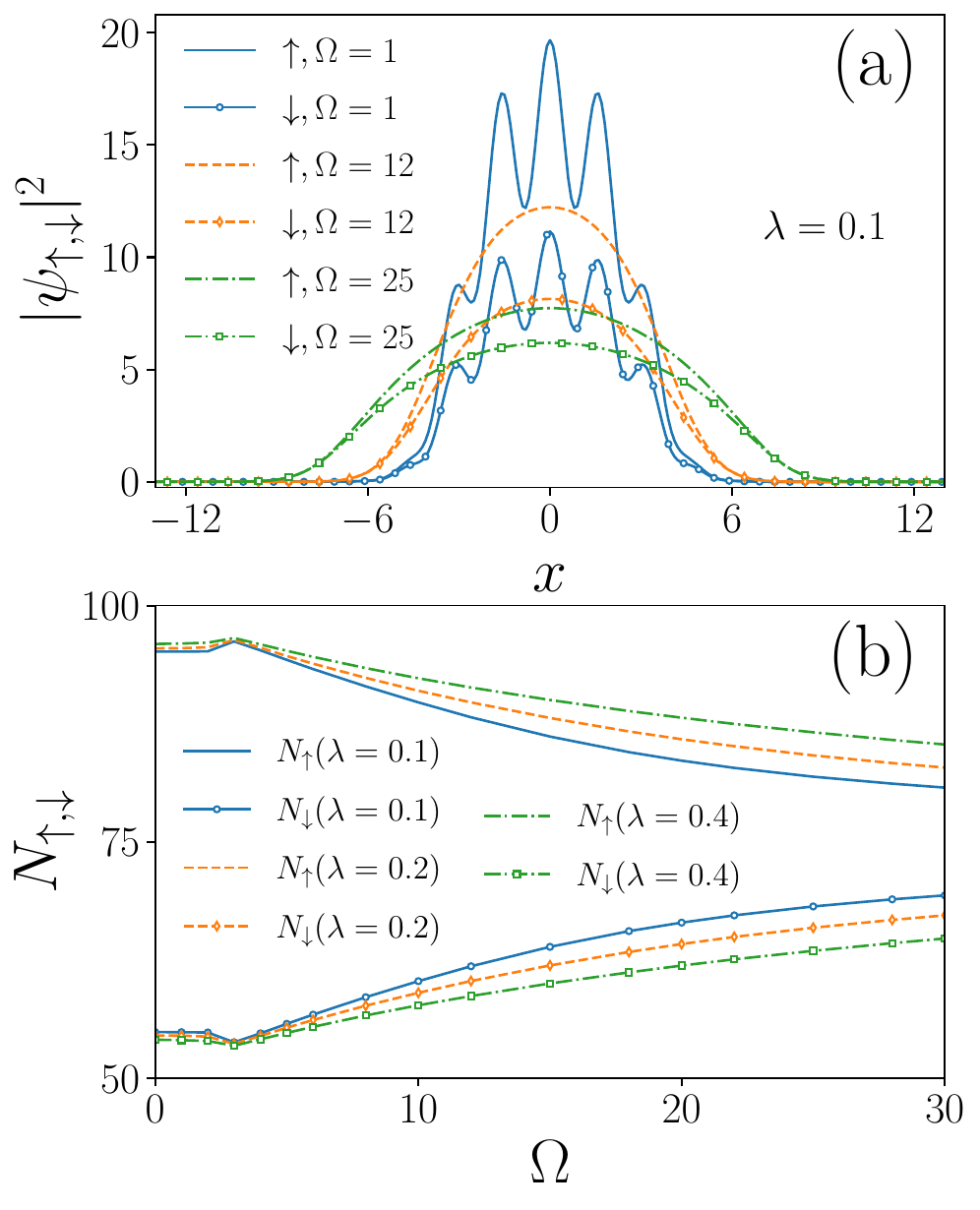}
\caption{(a) Trapped ground state droplet components for several $\Omega$ (see legend) and fixed atom number $N=150$ and $\lambda=0.1$. 
A deformation from stripe to Gaussian type droplets occurs followed by broadening of the droplet width and peak density reduction. 
(b) Atom number in each spin state for varying $\Omega$ and distinct $\lambda$ (see legend). 
The intracomponent interaction ratio is   $g_{\downarrow \downarrow}/g_{\uparrow \uparrow}=4$ with $g_{\uparrow \uparrow}=0.5$ and $\delta g=0.1$, the SOC wavenumber $k_L=2$.}
\label{fig:varyome1}
\end{figure}

The droplet-to-gas transition quantified through $\mu$ can be also clearly seen for different interaction imbalances that we have checked, see e.g.  Fig.~\ref{fig:denqdpwWithTrap}(c). 
In particular, the stronger the imbalance the larger the negative $\mu$ interval. 
Otherwise stated, for stronger intracomponent interactions the droplet character is sustained for a broader atom number which is important especially for tighter traps. The critical number of atoms, $N_c$, where the transition point occurs is explicitly shown in Fig.~\ref{fig:denqdpwWithTrap}(c) for varying $\lambda$ and specific interaction ratios verifying our previous arguments. 
Indeed, we find that $N_c$ increases for larger $g_{\downarrow\downarrow}/g_{\uparrow\uparrow}$, while it decreases with $\lambda$ in a power-law fashion with an exponent close to -1. 

\begin{figure}[!htp]
\centering\includegraphics[width=0.99\linewidth]{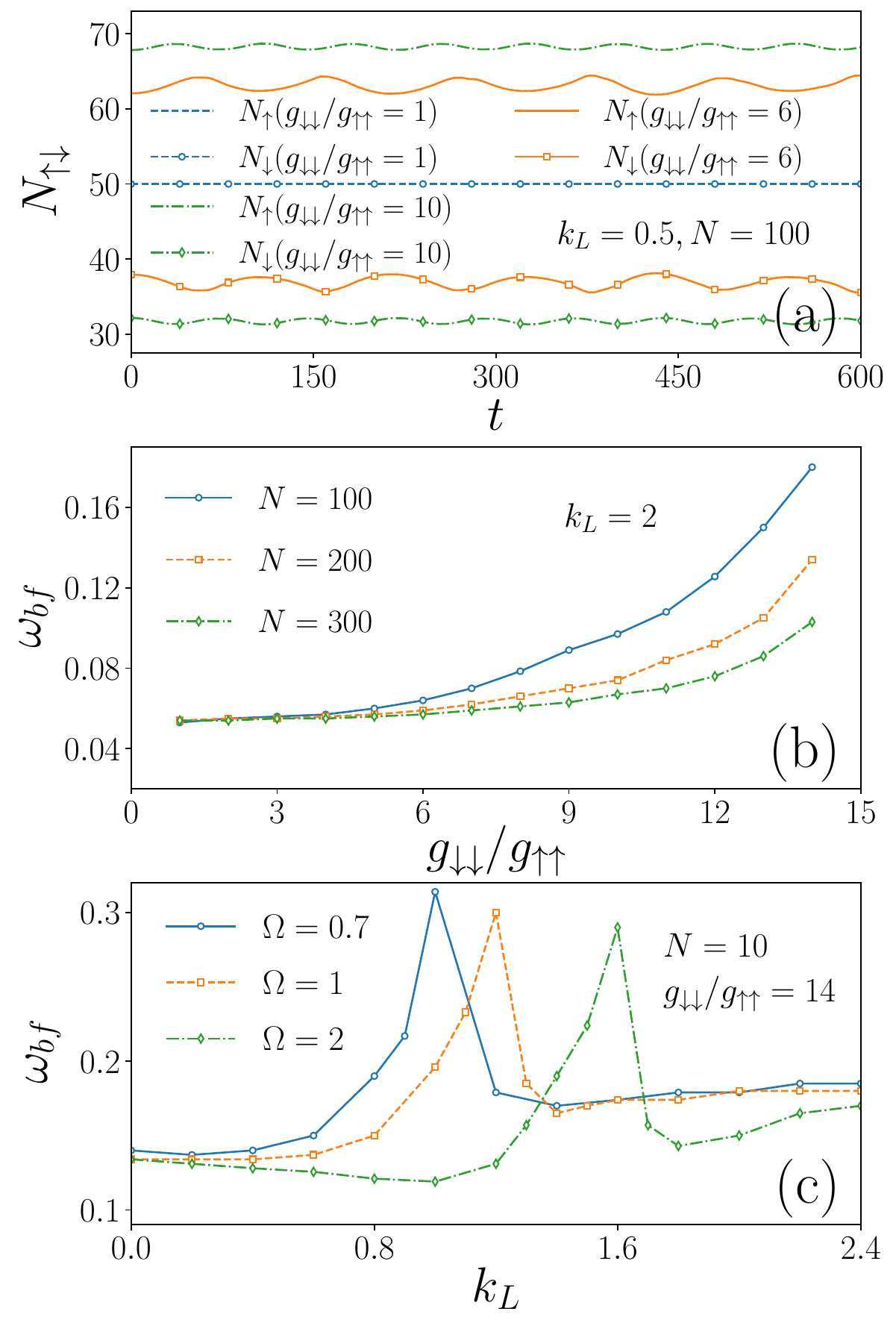}
\caption{(a) Time-evolution of the spin populations of the droplet for distinct $g_{\downarrow\downarrow}/g_{\uparrow\uparrow}$ (see legend) in the course of the collective breathing dynamics of the droplets with fixed $k_L=0.5$ and $\Omega=0.5$. 
(b) Breathing mode frequency, $\omega_{bf}$ of the droplet as a function of $g_{\downarrow \downarrow}/g_{\uparrow \uparrow}$ for (b) $k_L=2, \Omega=0.5$ and (c) different SOC wavenumbers $k_L$ at various $\Omega$ (see legend). 
The almost constant $\omega_{bf}$ in panel (b) refers to flat-top structures transforming into Gaussian droplets whose $\omega_{bf}$ increases with $g_{\downarrow\downarrow}/g_{\uparrow\uparrow}$. 
Also, at threshold from standard to stripe droplets, $\omega_{bf}$ is maximized in terms of $k_L$. 
In all the panels the initial droplet state is characterized by $g_{\uparrow \uparrow} =0.2$ and $\delta g=0.1$. 
The dynamics is triggered following a quench of the trap strength from $\lambda=0.01$ to $\lambda=0.03$ in panel (a,b) and $\lambda=0.01$ to $\lambda=0.05$ in panel (c).}
\label{fig:brefre}
\end{figure}

Turning to the impact of the Rabi-coupling on the ground state phases of the trapped droplets, we observe that it is the same as in free space, see Fig.~\ref{fig:varyome1}. 
Specifically, Fig.~\ref{fig:varyome1}(a) presents the droplet density profiles for different $\Omega$ and fixed  $g_{\downarrow\downarrow}/g_{\uparrow\uparrow}=4$, $N=150$ and  $\lambda=0.1$. 
The size (width) of the droplets increases while their peak densities decrease for larger $\Omega$. Interestingly, the droplet components also transform from stripe distributions for $\Omega=1$ to standard (i.e. non-modulated) droplet structures for increasing $\Omega$ where $k_L^2<\Omega$ holds. 
Yet also in this case, flat-top signatures are diminished when compared to the un-trapped case. It is, however, worth noting that the total chemical potential remains negative independently of $\Omega$ and, in particular, becomes more negative for larger $\Omega$, verifying the bound nature of the system. 
Of course, for a fixed $\Omega$ and varying $\lambda$, $\mu$ features an increasing trend towards positive values. 
Moreover, the population difference among the components becomes gradually smaller for increasing $\Omega$ and depends on $\lambda$, see also Fig.~\ref{fig:varyome1}(b).  
Namely, the spin balance process with $\Omega$ is delayed for larger $\lambda$.

\begin{figure*}[!htp]
\centering\includegraphics[width=0.99\linewidth]{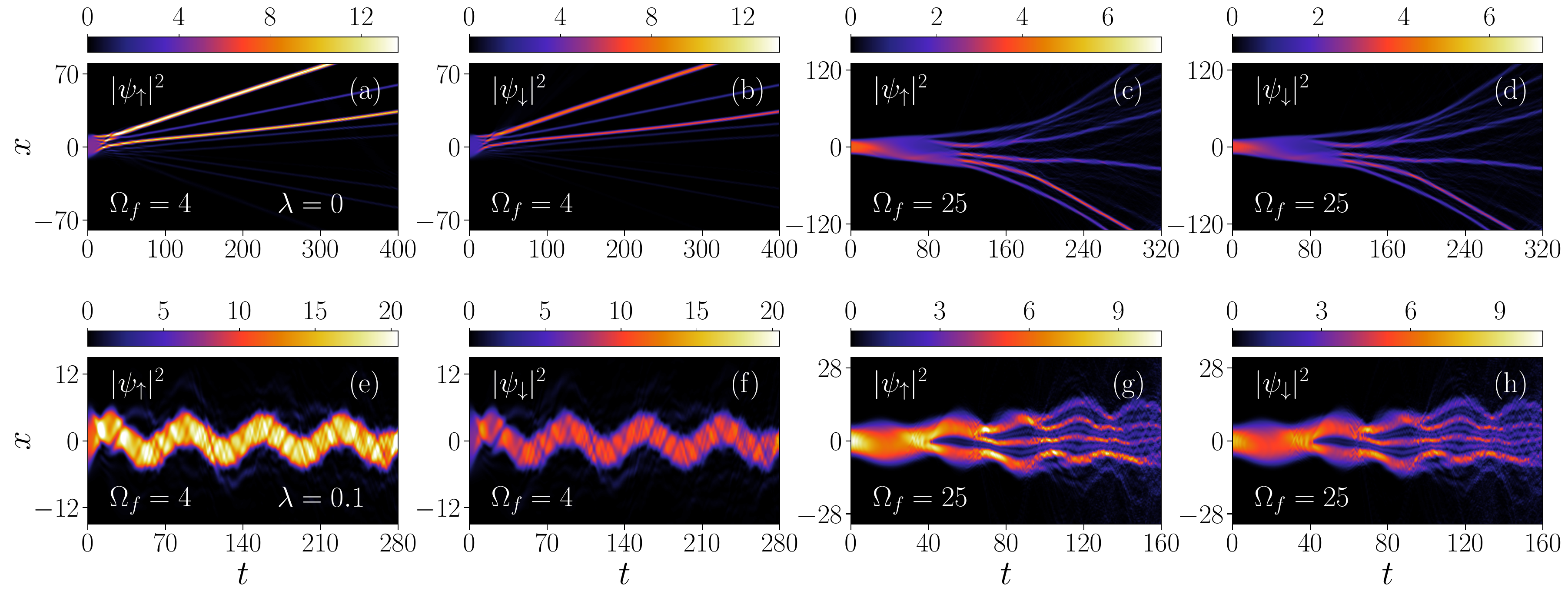}
\caption{Dynamical evolution of each component SOC droplet density after quenching the Rabi-coupling from $\Omega_i=15$ to $\Omega_f=4$ in the (a), (b) absence of a trap and (e), (f) presence of a trap with strength $\lambda=0.1$. 
The same as above but for $\Omega_i=15$ to $\Omega_f=25$ (c), (d) without and (g), (h) with the trap having $\lambda=0.1$. 
In the absence of the trap, a quench on $\Omega_f$ towards lower values results in the formation of fragmented moving droplets (panels (a), (b)), while quenches to higher $\Omega_f$ lead to erratic fragmented patterns (panels (c), (d)). 
The postquench state in the former (latter) quench is  dynamically (energetically) unstable as was shown in Ref.~\cite{Gangwar2024}. 
In the trap the SOC droplets oscillate with frequency equal to the one of the confinement. 
The ground state of the SOC droplet is prepared for $g_{\downarrow \downarrow}/g_{\uparrow \uparrow}=4$, $g_{\uparrow \uparrow}=0.5$, $\delta g=0.1$, $k_L=2$ and $N=150$.}
\label{fig:quenome}
\end{figure*}


\section{Quench dynamics of the imbalanced SOC droplets}\label{Dyn_drops}

\subsection{Collective breathing mode }
\label{sec:3c}

To develop a preliminary understanding of the dynamical response of imbalanced SOC droplets, we examine the properties of one of their fundamental collective excitations, namely the breathing mode. 
To generate this collective mode, we prepare the two-component droplet configuration in its ground state under the influence of a weak trap strength $\lambda=0.01$ while fixing the SOC parameters $\Omega=0.5$ and $k_L=0.5$. Subsequently, a quench of the trap strength to $\lambda=0.03$ is applied, inducing an overall expansion and contraction dynamics of each droplet component distribution manifesting the ensuing breathing mode evolution. 
Notably, the latter is accompanied by an underlying spin-mixing dynamics among the components as captured through $N_{\uparrow}(t)-N_{\downarrow}(t)$.  
This process is traced back to the existence of SOC (and, in fact, the Rabi-coupling) and can be easily deduced by monitoring the instantaneous population of each spin state depicted in Fig.~\ref{fig:brefre}(a). 
The component populations undergo an out-of-phase oscillatory behaviour, while both the oscillation amplitude and frequency depend on the interactions. 
We find that at weak interactions (e.g. $g_{\downarrow \downarrow}/g_{\uparrow \uparrow}=1$), the spin-mixing is negligible and becomes more pronounced at intermediate interactions (e.g. $g_{\downarrow \downarrow}/g_{\uparrow \uparrow}=6$) while it again reduces for stronger interactions (e.g. $g_{\downarrow \downarrow}/g_{\uparrow \uparrow}=10$). 
Moreover, it is possible to associate the aforementioned out-of-phase oscillatory behaviour of the populations with the different stages of breathing evolution. In particular, during the expansion interval, atoms get transferred from the spin-down to the spin-up component and vice versa in the course of the contraction stage. In all cases, independently of $g_{\downarrow \downarrow}/g_{\uparrow \uparrow}$ or the SOC parameters, the total atom number is conserved.

The interaction-dependent behaviour of the breathing mode frequency, $\omega_{bf}$, is explicitly provided in Fig.~\ref{fig:brefre}(b) for different atom numbers and constant SOC parameters $\Omega=0.5$ and $k_L=2$. 
Notice that $\omega_{bf}$ is calculated through the spectrum of the time-evolved droplet width of each component, namely $\braket{\psi_{\downarrow, \uparrow}|x^2|\psi_{\downarrow, \uparrow}}$~\cite{mistakidis2021formation} which can be experimentally assessed through in-situ absorption images~\cite{fukuhara2013quantum}. For the analysis below, $\omega_{bf}$ is the same for both components since, within the parameter regime considered, the droplets are not strongly particle imbalanced.    
Specifically, $\omega_{bf}$ remains almost constant around $\sim 0.05$ for all parameter sets  ($N$, $g_{\downarrow \downarrow}/g_{\uparrow \uparrow}$) which correspond to flat-top droplet distributions. However, beyond the interaction threshold where the droplets acquire a Gaussian type distribution, $\omega_{bf}$ starts to feature an increasing tendency. 
This increasing behaviour of $\omega_{bf}$ is caused by the substantially shrunk droplet distributions for larger $g_{\downarrow \downarrow}/g_{\uparrow \uparrow}$ [see Fig.~\ref{fig:denqdsw}(a)], which lead to a reduced period of the ensuing breathing evolution.  
A similar phenomenology of $\omega_{bf}$, in terms of $g_{\downarrow \downarrow}/g_{\uparrow \uparrow}$, occurs also for different SOC wavenumbers that we have checked, e.g. also for $k_L=0.5$ where standard non-modulated droplets exist. 
Recall that such an increasing tendency of $\omega_{bf}$ with respect to the interactions has also been reported in the case of a symmetric droplet~\cite{Gangwar2022}.

To reveal the interplay of the SOC parameters on $\omega_{bf}$ we next examine how it depends on $k_L$ variations for distinct $\Omega$ and fixed $g_{\downarrow\downarrow}/g_{\uparrow\uparrow}=14$ and $N=10$, see Fig.~\ref{fig:brefre}(c). 
It can be seen that at a given Rabi-coupling $\omega_{bf}$ exhibits an increasing trend reaching a maximum at SOC wavenumbers for which $\Omega\lesssim k_L^2$. 
Recall that this refers to the threshold set by the SOC parameters where the droplet transits from the standard to a stripe configuration. 
Indeed, for $k_L^2 >\Omega$, the breathing frequency decreases until it attains an almost saturated value of around $\omega_{bf}\sim 0.15$ for all $\Omega$ at sufficiently large $k_L$. 
We remark here that such a behavior of $\omega_{bf}$ in terms of $k_L$ at fixed $\Omega$ has also been discussed in the case of symmetric droplets~\cite{Gangwar2024}.

\subsection{Droplets fragmentation by  Rabi-coupling quenches}\label{sec:3d}

Next, we examine the dynamics of the SOC two-component droplet subjected to sudden changes of the Rabi-coupling with and without an external trap, see Fig.~\ref{fig:quenome}. 
In particular, we start with the ground state of the SOC droplets characterized by $\Omega=15$, $g_{\downarrow\downarrow}/g_{\uparrow\uparrow}=4, g_{\uparrow\uparrow}=0.5, k_L=2 $, $N=150$ and quench to either  $\Omega_f=4$ or $\Omega_f=25$. 
On the ground state level, it is known that in the former case of $\Omega_f=4$, the droplets have the tendency to form stripes since the Rabi-coupling becomes comparable to the SOC wavenumber and their population imbalance may increase [Fig.~\ref{fig:denqdpwithome} and Fig.~\ref{fig:varyome1}]. However, for $\Omega_f=25$ it is anticipated that in their ground state, the droplets will acquire a flat-top non-modulated shape with vanishing intercomponent atom imbalance, see also Fig.~\ref{fig:denqdpwithome} and Fig.~\ref{fig:varyome1}.

\begin{figure}[!ht]
\centering\includegraphics[width=0.99\linewidth]{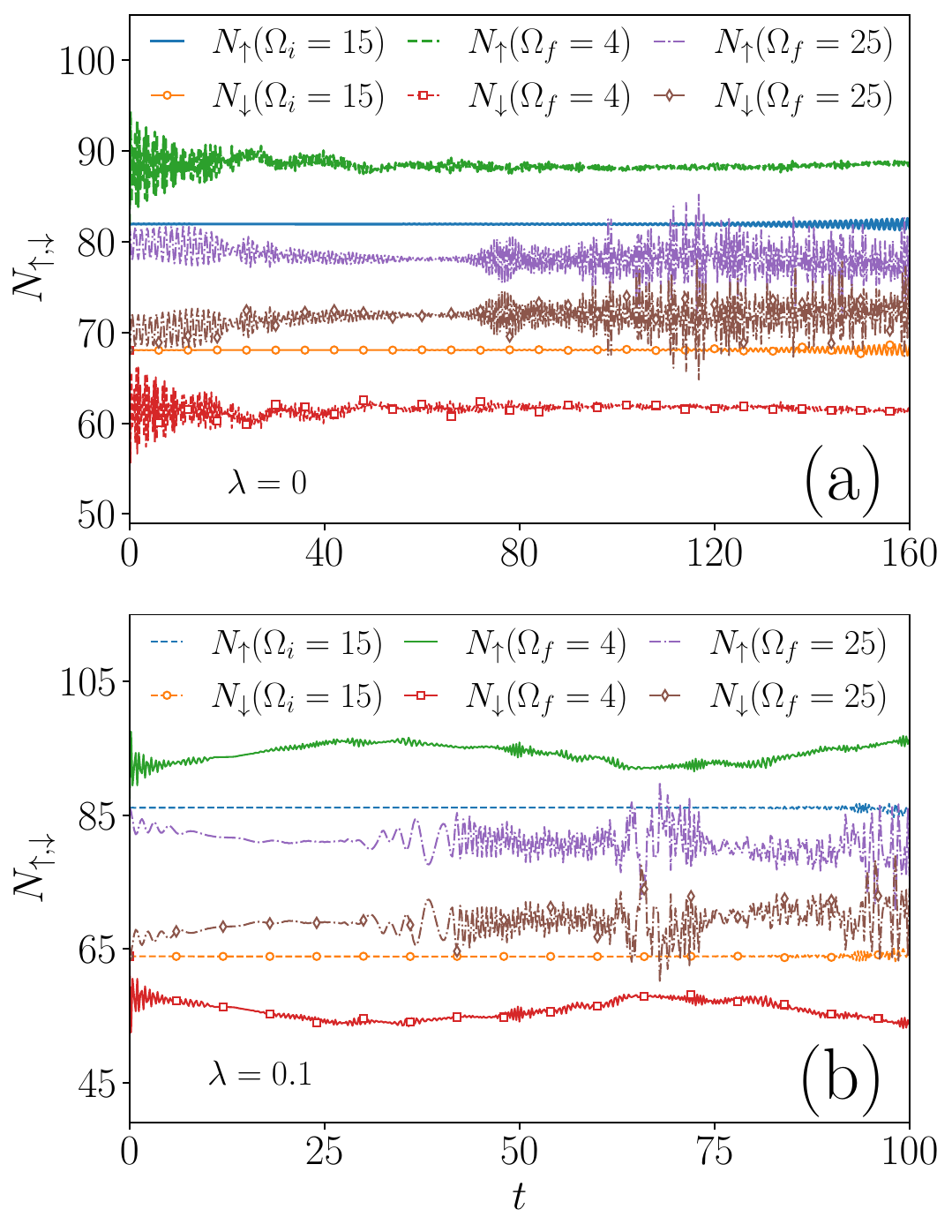}
\caption{Time-evolution of the SOC droplet spin populations upon considering different quenches of the  Rabi-coupling (see legends) (a) with and (b) without an external harmonic trap of strength $\lambda=0.1$. Despite the fluctuations in the individual atom numbers the populations do not feature significant exchange. }
\label{fig:varynorm}
\end{figure}

For quenches to $\Omega_f=4$, we observe that in the absence of a trap both SOC droplet components acquire the same finite velocity towards positive $x$ and fragment into several parts as illustrated Fig.~\ref{fig:quenome}(a), (b). 
As argued above, the fragmentation process is traced back to the ground state density favoring the build-up of stripe structures on top of the background which is dynamically unstable.
Interestingly, both components retain their initial miscible character in contrast to the expectations on the post-quench ground state level. 
Moreover, intercomponent atom exchange is evident in the spin populations shown in Fig.~\ref{fig:varynorm}(a) which from $N_{\uparrow} \approx 82$ and $N_{\downarrow} \approx 68$ fluctuates around $N_{\uparrow} \approx 89$ and $N_{\downarrow} \approx 61$ during the evolution. 
In the presence of the trap, the time evolution of the SOC droplets is altered. 
Indeed, both components after the quench gain the same finite velocity (as in the un-trapped scenario), they reach the trap edge and subsequently undergo an oscillatory motion around $x=0$ dictated by the trap frequency, see Fig.~\ref{fig:quenome}(e), (f). 
Notice also the pronounced density disturbances of both spin states which essentially reflect the excited nature of the droplets due to the quench. 
The spin transfer is once again relatively weak as shown in Fig.~\ref{fig:varynorm}(b), while performing an overall oscillatory behavior on top of the vanishing amplitude oscillations which are an imprint of the excited droplet character. 
Finally, once again and in contrast to the ground state behavior the droplets remain miscible in the course of the dynamics. 


On the other hand, a quench to larger $\Omega_f$, e.g. $\Omega_f=25$, leads to droplets breaking into different left and right moving ones in an erratic manner, see Fig.~\ref{fig:quenome}(c), (d). This behavior is attributed to the most probably energetically  unstable character of the  postquench state. For this reason the droplet breaks erratically as the time evolves allowing the system to reach a lower energy state. Specifically, at short evolution times ($t\lesssim 30$) the droplets take a velocity towards the negative $x$ direction accompanied by an asymmetric with respect to $x=0$ broadening of their profiles ($30\lesssim t\lesssim 90$) and consecutive fragmentation into smaller sized ones travelling in both directions as time evolves. 
As above, the components are miscible throughout the dynamics and their populations are only slightly changed tending to become almost equal at longer evolution times as depicted in Fig.~\ref{fig:varynorm}(a). 
The highly oscillatory small amplitude behavior of $N_{\uparrow}(t)$ and $N_{\downarrow}(t)$ reflects the excited nature of the droplets. 
For such quenches to larger Rabi-couplings, a similar droplet response takes place also upon considering an external trap, see Fig.~\ref{fig:quenome}(g), (h).    
However, in this case, the droplet components initially ($t\lesssim 40$) tend to oscillate within the trap and afterwards fragment as in the previous case. 
Naturally, the dispersion trend of the fragments is less compared to the un-trapped case since the trap sustains the droplet background. As a consequence, a more prominent interference occurs among the fragments since they lie close to one another. 
The quench-induced spin transfer is relatively small and at long evolution times, $t\gtrsim 90$, the components tend to acquire the same population. 
Also, miscibility is observed between the components in the course of the evolution as anticipated from the postquench ground state.

\subsection{Spin-demixing and counterpropagating droplets after quenching the SOC wavenumber}

\begin{figure}[!ht]
\centering\includegraphics[width=0.99\linewidth]{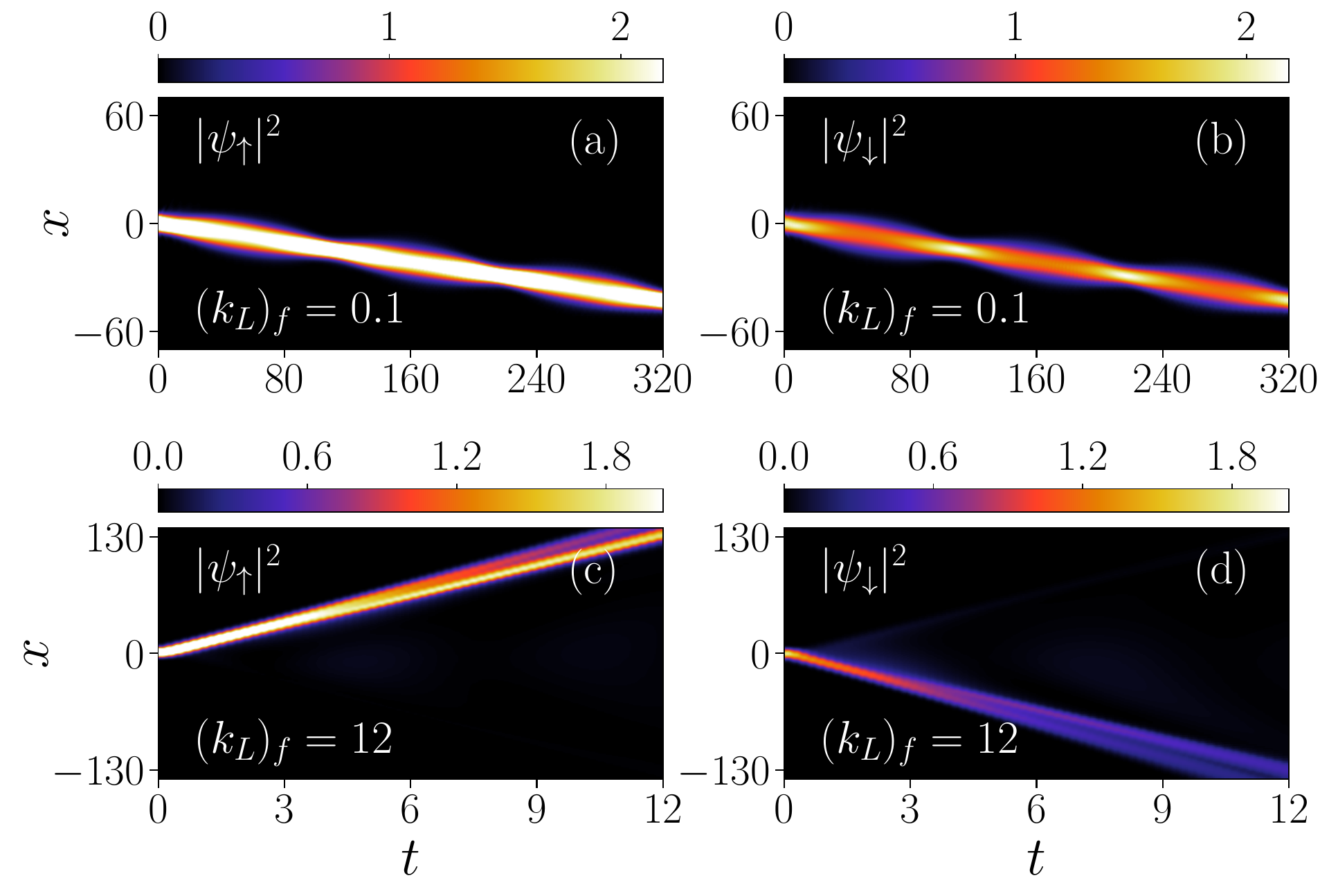}
\caption{Temporal evolution of each droplet component (see legends) density profile after quenching the SOC wavenumber from  $(k_L)_i=0.5$ to (a), (b) $(k_L)_f=0.1$ and (c), (d) $(k_L)_f=12$. 
It is observed that for large postquench SOC wavenumbers such that $(k_L)^2_f>\Omega$ [panels (c), (d)] the components demix during the evolution and move in opposite directions. Otherwise, the components are miscible and head towards the same direction.  
The two-component Bosonic system is prepared in its ground state with $g_{\downarrow \downarrow}/g_{\uparrow \uparrow}=10$, $g_{\uparrow \uparrow}=0.2$, $\delta g=0.1$, $k_L=0.5$, $\Omega=0.5$ and $N=50$.}
\label{fig:quenkl}
\end{figure}

Let us now turn our attention to the impact of the SOC wavenumber on the dynamical response of the two-component interaction imbalanced droplets both in the presence and absence of an external harmonic trap. 
Below, the droplet setting is initiated in each two-component ground state with parameters $g_{\downarrow \downarrow}/g_{\uparrow \uparrow}=10$,  $\delta g=0.1$, $k_L=0.5$, $\Omega=0.5$ and $N=50$. 
Focusing on the un-trapped case, we show the density evolution of each droplet component upon considering a quench to relatively small [Fig.~\ref{fig:quenkl}(a), (b)] or large [Fig.~\ref{fig:quenkl}(c), (d)] SOC wavenumbers quantified by $(k_L)^2_f<\Omega$ and  $(k_L)^2_f>\Omega$ respectively. 
Independently of the final value of the SOC wavenumber, each droplet component experiences a momentum ``kick" since $k_L$ is related to the system's momentum~\cite{zhai2015degenerate}, and hence the droplets become moving.   
In particular, in the case of $(k_L)^2_f<\Omega$ depicted in Fig.~\ref{fig:quenkl}(a), (b) we observe that both components feature a breathing oscillation while moving in the negative $x$-direction with a constant velocity. 
As such, we conclude that the droplet components remain mixed during the course of evolution. 
We also mention in passing that a relatively weak periodic population transfer among the components of maximum amplitude $\sim 6\%$ takes place (not shown) following the aforementioned droplets breathing.  

\begin{figure}[!ht]
\centering\includegraphics[width=0.99\linewidth]{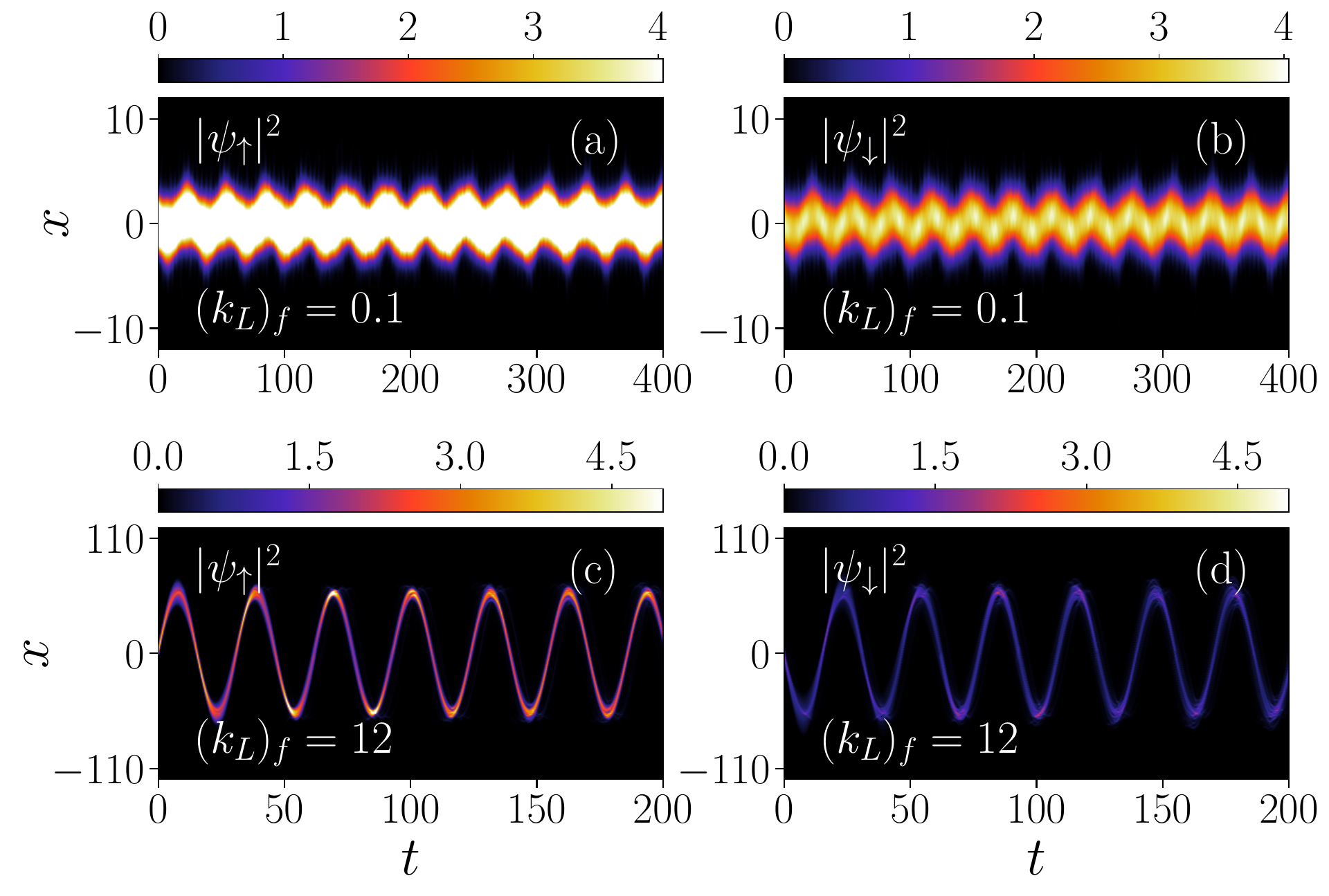}
\caption{Density dynamics of the different harmonically trapped droplet components following a sudden change of the SOC wavenumber from $(k_L)_i=0.5$ to (a), (b) $(k_L)_f=0.1$ and (c), (d) $(k_L)_f=12$. 
As it can be seen, for postquench values satisfying $(k_L)^2_f>\Omega$ [panels (c), (d)] the droplets oscillate out-of-phase within the trap; otherwise they oscillate in-phase.  
The imbalanced droplet setting is initialized in its  ground state configuration where $\lambda=0.2$, $g_{\downarrow \downarrow}/g_{\uparrow \uparrow}=10$, $g_{\uparrow \uparrow}=0.2$, $\delta g=0.1$, $\Omega=0.5$ and $N=50$.}
\label{fig:quenkl1}
\end{figure}


In contrast, quenching to $(k_L)_f^2>\Omega$ leads to a different droplet arrangement. Indeed, as illustrated in Fig.~\ref{fig:quenkl}(c), (d), the droplet components become counterpropagating and show a gradual spatial delocalization tendency during the evolution. 
This delocalization trend is reminiscent of the droplet fragmentation in the case of a symmetric mixture~\cite{Gangwar2022, Gangwar2024}. The fact that the droplets move oppositely is attributed to the dominant SOC contribution here, which appears with opposite signs in the eGPEs~(\ref{eq:gpsoc:2}), imparting opposite velocities to the different components.  
Evidently, the components are demixed since their overlap is vanishing throughout the propagation, while there is a small periodic population exchange accompanying the dynamics (not shown).

The existence of an external harmonic trap does not substantially alter the droplet response in terms of spin-demixing but rather induces an oscillatory droplet motion, see Fig.~\ref{fig:quenkl1}. 
For completeness, again, we consider quenches, both to smaller and larger SOC wavenumbers, while the remaining system parameters remain the same as in the un-trapped scenario. 
An overall prevailing feature is that the specific quench triggers a directed droplet in-trap oscillatory motion, which depends further on the postquench $k_L$ value. 
For instance, when $(k_L)_f=0.1<\sqrt{\Omega}$ the droplet components oscillate in sync with the same frequency as the trap frequency [Fig.~\ref{fig:quenkl1}(a), (b)]. They also retain their miscible character during the entire evolution, and intercomponent atom transfer is suppressed. 
However, for $(k_L)^2_f=12>\Omega$, the droplet components exhibit out-of-phase oscillations [Fig.~\ref{fig:quenkl1}(c), (d)] and therefore become immiscible during the dynamics. 
The oscillation period is dictated by the external trap, whilst the spin exchange population is relatively weak of maximum amplitude $\sim 15\%$ and not regular. 
We remark that a similar spin-mixing and spin-demixing dynamics was reported for the SOC solitons after the sudden introduction of SOC into the system~\cite{Ravisankar2020}.

\section{Summary and Perspectives}
\label{sec:4}

We have investigated the ground state phases and the quench dynamics of interaction-imbalanced SOC droplets in one dimension. 
Our analysis is based on the suitable coupled system of extended Gross-Pitaevskii equations of motion in the presence and absence of an external harmonic trap. 
Particularly, a variety of different ground state droplet configurations are identified that 
host miscible components, and range from standard non-modulated to stripe patterns featuring flat-top or Gaussian density profiles. 
A key feature of our setup is the presence of SOC, which through the Rabi-coupling favors population transfer between the involved spin states and thus atom number imbalance. 

It is found that droplets in both components deform from flat-top to Gaussian ones for increasing (decreasing) intracomponent interaction ratios (total atom number). 
On the other hand, the stripe (spatially modulated) droplet character is regulated by the SOC wavenumber. 
Spin imbalance increases with larger interaction strength, higher SOC wavenumber, or smaller Rabi coupling. 
Interestingly, an increase of the Rabi-coupling leads gradually to  structural modifications from stripe to non-modulated Gaussian droplets and subsequently to flat-top ones whose imbalance is suppressed. This transition is also evident in the spin populations acquiring an almost constant imbalance in the stripe region and start decreasing in the Gaussian regime while tending to be almost the same when approaching the flat-top regime.


On the other hand, in the presence of a trap, the stripe droplets feature  Gaussian profiles with their size and peak density increasing for larger atom number or decreasing interactions. 
Importantly, a transition from a self-bound to a trapped gas phase occurs for increasing atom number as captured by the total chemical potential which changes from negative to positive.  
The underlying critical region shifts to smaller atom numbers for increasing trap frequency or to a larger number of atoms for fixed trap frequency and stronger intracomponent repulsions.  
A similar phenomenology to the un-trapped case in terms of the Rabi-coupling takes place also in the presence of a trap but instead of reaching flat-top configurations Gaussian ones are attained.

Turning to the dynamics we have first examined the behavior of the collective breathing motion of the two-component droplet system after a quench of the trap strength. This process induces a relatively small population transfer. It is found that the breathing mode frequency increases for larger intracomponent interaction ratio, while it features a maximum in terms of the SOC wavenumber at the location of the transition from a standard non-modulated to the stripe droplet. 
Additionally, a sudden reduction or increase of the Rabi-coupling is associated with minor intercomponent population exchange and the formation of fragmented moving droplets in the absence of the trap and oscillating ones with the trap frequency in its presence. 
On the other hand, following a quench of the SOC wavenumber leads to a drastically different dynamical response. 
In particular, independently of the postquench SOC wavenumber the droplets acquire finite momentum and feature a directed motion. 
In the absence of a trap, this motion refers to droplet propagation in the same or opposite direction for small and large post-quench amplitudes respectively. 
This process is accompanied by spatial mixing or de-mixing of the droplet components which simultaneously exhibit breathing and spatial delocalization respectively. 
The presence of the trap enforces an oscillatory droplet motion characterized by the trap frequency. 
Out-of-phase (in-phase) droplet motion is evidenced for larger (weak) quench amplitudes wit the droplets being immiscible (miscible).  

Based on our results, there is a multitude of intriguing extensions that may be followed in the future. An immediate step is to construct an effective potential picture in the case of large particle imbalance where the majority stripe droplet component plays the role of an external lattice potential for the minority one as was done in Ref.~\cite{englezos2023particle} but in the absence of SOC. 
Another interesting pathway is to calculate the excitation spectrum of these two-component droplet structures, especially focusing on the stripe configurations and their crossover to non-modulated droplets.~\cite{Ravisankar2021b, sgangwar2024emergence, Gangwar2024}. Moreover, it would be worth examining the correlation patterns~\cite{mistakidis2021formation} associated with the stripe droplets, shedding light on the underlying two-body processes participating in the respective many-body states.  {An additional appealing   direction would be to generalize the considered model in order to create linear droplet arrays similarly to the recent work of Ref.~\cite{Zhang2019} in order to explore their transport or entanglement properties.}
The generalization of our setup in  two dimensions in order to reveal the phases of two-component stripe droplets and their relation with supersolids~\cite{chomaz2022dipolar} is desirable. 
Finally, it would be interesting to study the impact of optomechanical couplings~\cite{pradhan2024ring,pradhan2024cavity} in the identified droplet states.

\acknowledgments
S.G. would like to acknowledge the financial support from the University Grants Commission - Council of Scientific and Industrial Research (UGC-CSIR), India.  We also gratefully acknowledge our supercomputing facilities Param-Ishan and Param-Kamrupa (IITG), where all the numerical simulations were performed. 
The work of PM is supported by MoE RUSA 2.0 (Physical Sciences -- Bharathidasan University). 
S. I. M is supported from the Missouri University of Science and Technology, Department of Physics, Startup
fund. 
S. I. M acknowledges
fruitful discussions with P. G Kevrekidis, G. C. Katsimiga and G. Bougas on the topic of droplets. R.R. acknowledges the postdoctoral fellowship supported by Zhejiang Normal University, China, under Grants No. YS304023964.


\bibliography{references.bib}

\end{document}